\documentclass[a4paper]{article}

\usepackage{INTERSPEECH2022}

\usepackage{color}
\usepackage{enumitem}
\usepackage{hyperref}
\usepackage{multirow,makecell}
\usepackage{etoolbox,siunitx}
\robustify\bfseries
\usepackage{booktabs}
\usepackage{balance}
\usepackage{amssymb}
\usepackage{bm}
\usepackage{arydshln}
\usepackage{nicefrac}
\usepackage{comment}

\usepackage{cite}
\usepackage{bbm}
\usepackage[scr=boondoxo]{mathalfa}

\newcommand{\spaceintable}{1.4cm}

\def\R{\mathbb{R}}

\def\setC{\mathscr{C}}
\def\setD{\mathscr{D}}
\def\setV{\mathscr{V}}

\def\ct{v}
\def\cmap{\mathcal{C}}

\def\C{\mathcal{C}}
\def\D{\mathcal{D}}
\def\E{\mathcal{E}}
\def\G{\mathcal{G}}
\def\L{\mathcal{L}}
\def\S{\mathcal{S}}
\def\U{\mathcal{U}}
\def\one{\mathbbm{1}}

\usepackage{pifont}%
\let\oldding\ding%
\renewcommand{\ding}[2][1]{\scalebox{#1}{\oldding{#2}}}%

\def\w{{\bm{\theta}}}
\def\c{\mathbf{c}}
\def\mix{\mathbf{x}}
\def\es{\widehat{\mathbf{s}}}
\def\est{\widehat{\mathbf{s}}_{\operatorname{T}}}
\def\eso{\widehat{\mathbf{s}}_{\operatorname{O}}}
\def\s{\mathbf{s}}
\def\st{\mathbf{s}_{\operatorname{T}}}
\def\so{\mathbf{s}_{\operatorname{O}}}

\title{Heterogeneous Target Speech Separation}
\name{Efthymios Tzinis$^{1,2}$\thanks{Work performed as E. Tzinis was an intern at MERL. E. Tzinis and P. Smaragdis were partially funded by NIFA grant \#2020-67021-32799. Code and data recipes: \href{https://github.com/etzinis/heterogeneous_separation}{github.com/etzinis/heterogeneous\_separation}}, Gordon Wichern$^1$, Aswin Subramanian$^1$, Paris Smaragdis$^{2}$, Jonathan Le Roux$^1$} 
\address{
  $^1$Mitsubishi Electric Research Laboratories (MERL), Cambridge, MA, USA\\
$^{2}$University of Illinois at Urbana-Champaign, Urbana, IL, USA}
\email{\{etzinis2,paris\}@illinois.edu,\{wichern,subramanian,leroux\}@merl.com}

\begin{document}

\maketitle
\setlength{\abovedisplayskip}{3pt}
\setlength{\belowdisplayskip}{3pt}
\begin{abstract}
  We introduce a new paradigm for single-channel target source separation where the sources of interest can be distinguished using non-mutually exclusive concepts (e.g., loudness, gender, language, spatial location, etc). Our proposed heterogeneous separation framework can seamlessly leverage datasets with large distribution shifts and learn cross-domain representations under a variety of concepts used as conditioning. Our experiments show that training separation models with heterogeneous conditions facilitates the generalization to new concepts with unseen out-of-domain data while also performing substantially higher than single-domain specialist models. Notably, such training leads to more robust learning of new harder source separation discriminative concepts and can yield improvements over permutation invariant training with oracle source selection. We analyze the intrinsic behavior of source separation training with heterogeneous metadata and propose ways to alleviate emerging problems with challenging separation conditions. We release the collection of preparation recipes for all datasets used to further promote research towards this challenging task.
\end{abstract}
\noindent\textbf{Index Terms}: target speech separation, semi-supervised learning, heterogeneous conditions, conditional inference
\section{Introduction}

As exemplified in the cocktail party problem~\cite{cherry1953some}, humans have the uncanny ability to %
focus on a source of interest within a complex acoustic scene, and may change the target of their focus depending on the situation, %
relying on attention mechanisms that modulate the cortical responses to auditory stimuli~\cite{fritz2007auditory,mesgarani2012selectiveCorticalRepresentation}. 
While the field of source separation has made great strides towards reproducing such abilities in machines, particularly with the advent of deep learning approaches, there is still a gap in terms of the flexibility with which the target source can be determined.

Early works developed ``specialist'' models intended to isolate %
only a particular type of sound, such as for speech enhancement \cite{xu2014experimental,Weninger2014RNN,erdogan2015psf,wang2018supervised} or  instrument demixing \cite{jansson2017singing}, %
where the target was determined by the training scheme and could not be changed at test time. %
Later works such as deep clustering and permutation invariant training (PIT) \cite{hershey2016deepclustering,Isik2016Interspeech09,Yu2017PIT} focused on separating all sources in a mixture without an a priori differentiating factor. However, this lack of explicit bias can lead to %
instability~\cite{yousefi2019probabilistic,huang2021stabilizing}, 
and re-introducing some bias, for example via fixed assignments based on energy or speaker ID~\cite{yang2020interrupted}, may help stabilize training. %

Conditioned models, in which the target source of a system is determined based on some semantic information given as input, either via a sound class (e.g., speaker ID, instrument type) or an exemplar (e.g., reference utterance by a target speaker), %
can also benefit from the good training properties of an explicit bias while allowing some flexibility in choosing the target source at test time. Such methods %
can provide significant gains over PIT for speech \cite{delcroix2018single,ochiai2019unifiedInformedSpeakerExtraction,wang2019voicefilter,xiao2019single, zhuo_loc}, music \cite{Seetharaman2019ICASSPclass,Meseguer19CUNet,slizovskaia2021cunet}, and universal sound source separation \cite{tzinis2020improving,ochiai2020listen,okamoto2021environmentalOnomatopoeia}, but conditioned models have so far each only considered a single type of condition looking at extracting distinct sources based on mutually-exclusive criteria.

Arguably, the problem of sound source separation is not as well-posed as the way it has been considered in the past. There are many ways to ``slice'' an acoustic scene depending on the application or the user's intention, where the grouping of sounds considered as target may vary. %
A first attempt in this direction was motivated by the fact that sounds can be considered as part of hierarchies, and a desired target may lie at various points along such a hierarchy \cite{Manilow2020ISMIR10}. 
Here, we explore another direction, by proposing to make the conditioned models mimic humans' flexibility when selecting which source to attend to, by focusing on extracting sounds based on semantic concepts and criteria of different nature, i.e., heterogeneous, such as whether a speaker is near or far from the microphone, being soft or loud, or speaks in a certain language. The main contributions of this paper are:
\begin{figure}[t]
  \centering
  \includegraphics[trim={0.02cm 0.2cm 0.78cm 0.1cm}, clip,width=0.87\linewidth]{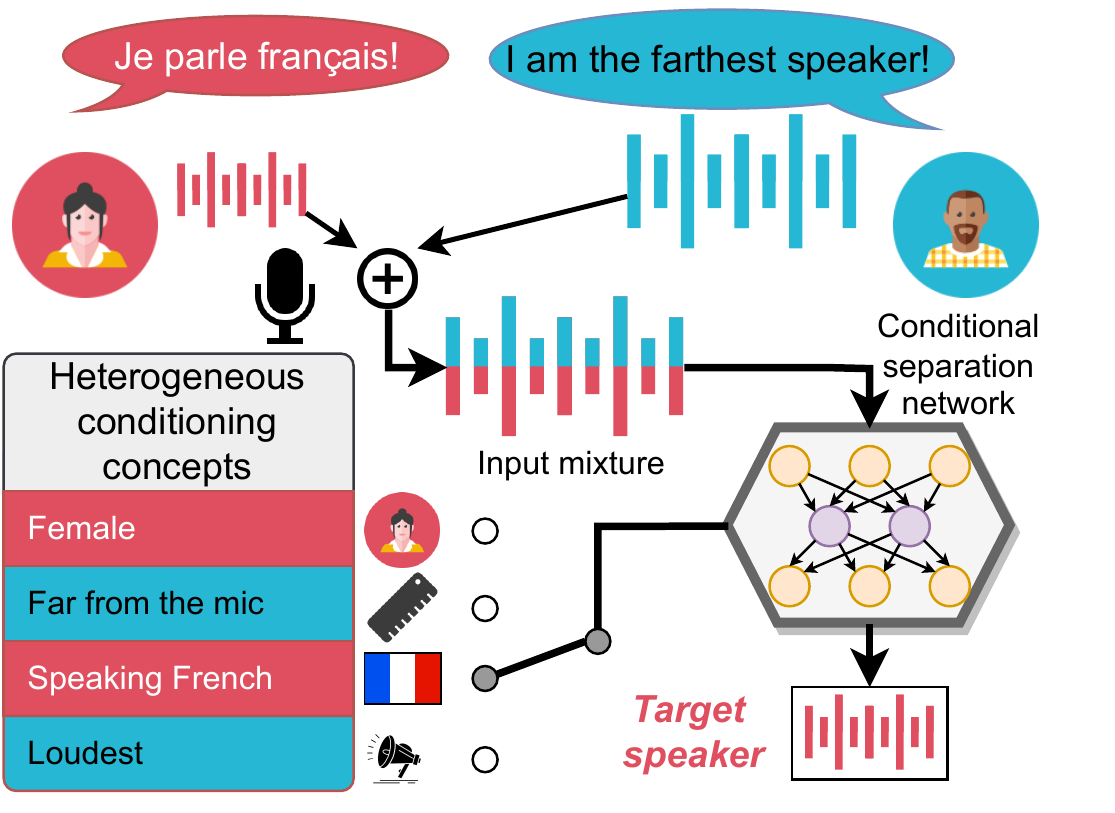}
  \vspace{-.2cm}
\caption{Illustration of the heterogeneous target speech separation task. Notice that speakers can be separated using any of the semantic concepts and speaker attributes on the left.}
\vspace{-.5cm}
\label{fig:heterogeneous_task}
\end{figure}
\begin{itemize}[leftmargin=*]
    \item We introduce a novel heterogeneous target source separation task and publicly release the associated datasets. 
    \item We propose a simple neural network architecture which can effectively separate target speech sources based on several non-mutually exclusive signal characteristic conditions, often outperforming PIT-based models with oracle %
    assignment.
    \item We make several experimental discoveries:
    1) heterogeneous conditioning can help cross-domain generalization, 2) 
    robustness to non-discriminative concepts can be achieved with a small amount of such examples without impacting the overall conditional performance, and
    3) adding extra ``easy'' conditions can lead to 
    better learning on more difficult conditions.
\end{itemize}

\section{Heterogeneous Target Source Separation}
\label{sec:method}

\subsection{Task formulation}
\label{sec:method:task}
We consider a mixture $\mix = \sum_{j=1}^{N} \s_j \in \R^{T}$ of $N$ source waveforms $\s_1,\dots,\s_N$, with $T$ time-domain samples. In general, we assume that there exists a signal characteristic condition $\C$ (e.g., the spatial location of a source) in a set $\setC$ of conditions, and a desired concept value $\ct$ for that condition (e.g., \textit{far} or \textit{near}) which belongs to the set $\setV$ of all discriminative concepts. Now, given the condition $\C$ and its concept value $\ct$, we would like to retrieve from the input mixture $\mix$ the \textit{target} submix $\st$ of all sources whose condition $\C$ matches the concept value $\ct$:
\begin{equation}
\label{eq:target}
    \st = {\textstyle \sum}_{j=1}^{N} \delta(\cmap(\s_j)=\ct) \s_j, 
\end{equation}
where $\delta$ is %
an indicator function, and we use the same notation $\C$ to denote a signal characteristic and the function $\cmap:\R^{T} \rightarrow \setV$ which returns the value of that characteristic for an input signal. %

In this work, as an illustration, we focus on the case of speech sources, and consider signal characteristics $\C$ in the set $\setC \!=\! \{\E, \G, \S, \L \}$, where $\E$ denotes the signal-energy (with values low/high), $\G$ the gender (female/male as self-identified by the dataset's speakers), $\S$ the spatial location (near/far), and $\L$ the language (English/French/German/Spanish). %
Thus, we can specify a target based on a total of $|\setV| = 2 + 2 + 2 + 4 = 10$ concepts. %
\begin{figure}[]
  \centering
  \includegraphics[trim={0.2cm 0.2cm 2.85cm 0.13cm}, clip,width=1.\linewidth]{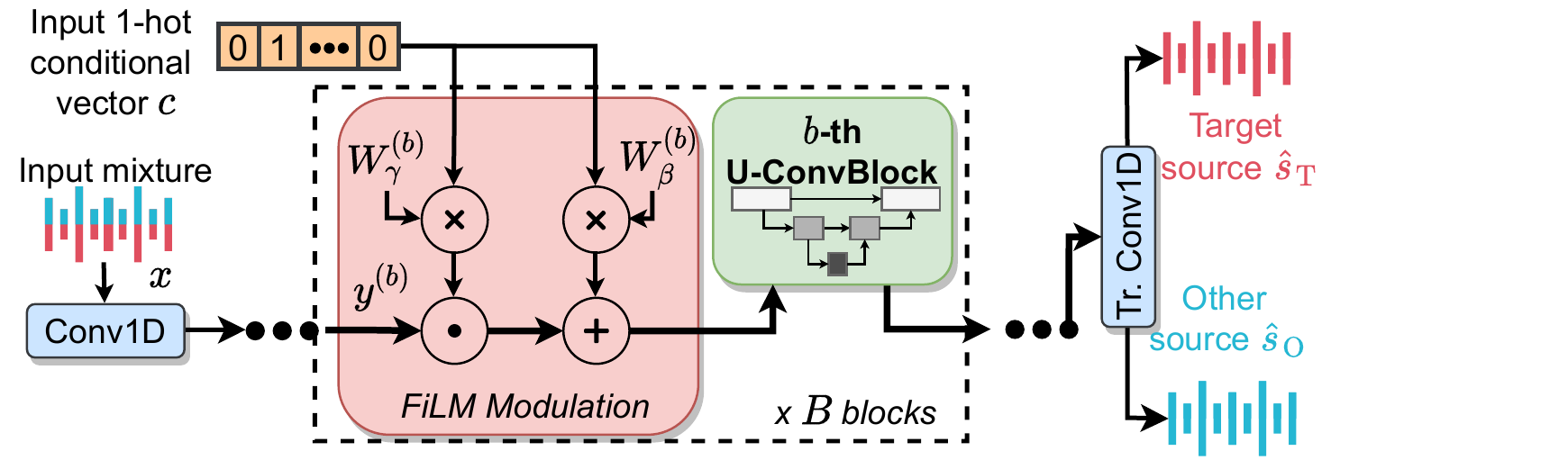}
  \vspace{-.6cm}
\caption{Illustration of the conditional Sudo rm -rf audio separation architecture. We show the FiLM modulation which is applied to the input $\mathbf{y}^{(b)}$ of the $b$-th U-ConvBlock in the stack.}
\vspace{-.5cm}
\label{fig:conditional_sudo}
\end{figure}
We can encode the semantic discriminative information for the desired concept $\ct$ in a one-hot vector $\c = \one[\ct] \in \{0, 1\}^{|\setV|}$ which has one only at the corresponding index of the concept $\ct$, given some arbitrary ordering of $\setV$. 
The goal of the task is then to train a separation model $f$, parameterized by $\w$, which takes as input a mixture of sources $\mix$ alongside a conditioning vector $\c$ and estimates the target submix $\est$ as follows:
\begin{equation}
\label{eq:cond_estimation}
    \est = f(\mix, \c; \w).
\end{equation}
\subsection{Proposed framework}
\label{sec:method:training}
We construct a new conditional separation model based on the efficient Sudo rm -rf default model configuration \cite{tzinis2022compute}, which obtains high fidelity audio reconstruction for source separation problems with minimal memory and computational requirements. Specifically, we add a FiLM \cite{film} modulation layer at the input of each of the $B$ U-ConvBlocks, as shown in Fig.~\ref{fig:conditional_sudo}. The extra parameters for the scaling and the bias are $B$ pairs of matrices ($\mathbf{W}_{\beta}, \mathbf{W}_{\gamma}$) with size $|\setV| \times C_{\operatorname{in}}$, where $C_{\operatorname{in}}=512$ is the number of intermediate channels in each processing block. Notably our proposed method, could theoretically scale up to an infinite amount of concepts since there is a dedicated $512$-dimensional parameterization for each one-hot representation.
We set the network $f$ to produce estimates $\est$ and $\eso$ for $\st$ and the submix $\so$ of \textit{other} (non-target) sources, enforcing $\est + \eso = \mix = \st + \so$ via a mixture consistency layer \cite{wisdom2019differentiableMixtureConsistency}.

We assume that we have access to a set of domains/datasets $\setD = \bigcup_j \mathcal{D}_j$, from where we sample clean sources and their semantic concepts to train our model by minimizing the $l_1$ loss:
\begin{equation}
\label{eq:loss}
    \begin{gathered}
    L_{\w} = |\est - \st| + |\eso - \so|, \enskip \est, \eso = f(\mix, \c; \w), \\
    \mix = \s_1 + \s_2, \enskip \s_i \sim \mathcal{D}, \enskip \mathcal{D} \in \setD, \enskip \c \sim P(\ct), \enskip \ct \in \setV,
    \end{gathered}
\end{equation}
where the target submix $\st$ is synthetically constructed as shown in Eq.~\ref{eq:target} given the sampled conditioning vector $\c$ and the sources $\s_i$. Notice that the sampling procedures for $\c$ and $\s_i$ are independent and thus, the same mixture $\mix$ can heterogeneously disentangled in multiple ways. We argue that controlling the sampling prior of a certain concept $P(\ct)$ is key to robust learning as well as out-of-domain generalization.  

\section{Experimental Framework}
\label{sec:exp_framework}

\subsection{Data collections}
\label{sec:exp_framework:datasets}
To show how heterogeneous conditional training behaves in various cross-domain training schemes with large distribution shifts (anechoic vs reverberant) as well as incorporate multiple conditions, we synthesize new conditioning datasets using Pyroomacoustics \cite{pyroomacoustics}. Generated mixtures are based on single-speaker utterances from the following three data collections.

\begin{table}[]
\scriptsize
\centering
\sisetup{table-number-alignment = center,detect-weight=true,detect-inline-weight=math}
\caption{Data collection metadata.}
\vspace{-.4cm}
\setlength{\tabcolsep}{0.8em}
\resizebox{.93\linewidth}{!}
{%
\begin{tabular}[t]{lccc}
\toprule
Metadata & WSJ & SLIB & SVOX \\
\midrule
Conditions $\setC $ & $ \{\E, \G\}$ &  $\{\E, \G, \S \}$ & $ \{\E,   \L, \S \}$ \\ 
\midrule
Room height (m) & - &  $ \U[2.6, 3.5] $ &  $ \U[2.75, 3.25] $ \\
Room length (m) & - &  $ \U[9.0, 11.0] $ &  $ \U[8.0, 10.0] $ \\
Room width (m) & - &  $ \U[9.0, 11.0] $ &  $ \U[8.0, 10.0] $ \\
RT 60 (sec) & - &  $ \U[0.3, 0.6] $ &  $ \U[0.4, 0.6] $ \\
Microphone location & - & Center & Center \\
Source height (m) & - &  $ \U[1.5, 2.0] $ &  $ \U[1.6, 1.9] $ \\
Far field distance (m) & - &  $ \U[1.7, 3.0] $ &  $ \U[1.5, 2.5] $ \\
Near field distance (m) & - &  $ \U[0.2, 0.6] $ &  $ \U[0.3, 0.5] $ \\
\midrule
 Number of test recordings  &$  1{,}770  $&$  2{,}620  $&$  11{,}083 $\\
 Number of test speakers  &$  18  $&$  40  $&$  294 $\\
 Number of train recordings  &$  8{,}769  $&$  132{,}553  $&$  124{,}937 $\\
 Number of train speakers  &$  101  $&$  1172  $&$  2347 $\\
 Number of val recordings  &$  3{,}557  $&$  2{,}703  $&$  10{,}244 $\\
 Number of val speakers  &$  101  $&$  40  $&$  279 $\\
\bottomrule
\end{tabular}
\label{table:datasets}
}\vspace{-.5cm}
\end{table}

\noindent \textbf{Wall Street Journal (WSJ)}: Anechoic English utterances from the WSJ0 corpus \cite{garofolo1993csr} following the same clean audio extraction and partitioning recipe as in the widely used WSJ0-2mix \cite{hershey2016deepclustering}.

\noindent \textbf{Spatial LibriSpeech (SLIB)}: Synthetically reverberant versions of $\{\textit{train-360}|\textit{dev}|\textit{test}\}$-\textit{clean} LibriSpeech~\cite{librispeech_dataset} partitions.

\noindent \textbf{Spatial LibriSpeech (SVOX)}: We use clean speech recordings available from the multi-lingual Voxforge \cite{voxforge} data collection. We use all available files under ``English'' ($53\%$), ``French'' ($15\%$), ``German'' ($16\%$), and ``Spanish'' ($16\%$). Since there is no specified partitioning, we randomly split the speakers at a ratio of $8\!:\!1\!:\!1$, for train, validation, and test, respectively. 

\subsection{Mixtures and conditions generation configurations}
\label{sec:exp_framework:mixture_generation}
All mixture datasets are synthetically generated on-the-fly (see the uniform sampling range for each parameter in Table \ref{table:datasets}), with fixed random seeds for the reproducibility of each partition. The exact data preparation recipes are available online. 

For each training epoch, we generate $20{,}000$ new on-the-fly mixtures consisting of two speakers, while for validation and testing we have a fixed-seed generation process of $3{,}000$ mixtures per separation concept $\ct \in \setV$. For each mixture, we sample a domain $\D \in \setD$ and a valid separation concept for the chosen domain $\D$, according to a specified prior $\ct \sim P(\ct)$. Then, we sample from $\D$ two sources which can be separated using the concept value $\ct$, and mix them with a uniformly sampled overlap between $[75, 100] \%$. For cross-domain training we always assume equal prior among the data collections and create only single-domain mixtures. In the case of in- or cross-domain training with the WSJ and SVOX datasets, we create mixtures with an input SNR in $\U[0, 5]$ and $\U[0, 2.5]$, respectively. The SLIB dataset is always used in cross-domain training with either WSJ or SVOX and follows the corresponding input SNR distribution. In order to magnify the domain mismatch for the WSJ-SLIB cross-domain training pair, we assume that SLIB mixtures always contain one near-field and one far-field speaker. On the contrary, when performing cross-domain training with SVOX, we assume that the prior of sampling near/near-field, far/far-field, and near/far-field utterance combinations is equal (except when conditioning on the spatial location $\S$, which needs the sources to be spatially separable, i.e., $\S(s_1) \neq \S(s_2)$). Consequently, we found that SVOX, which is also male-dominated, can become particularly challenging for the language conditioning, where all spatial location combinations are present.

Our method allows us to also create degenerate conditions where the constituent sources map to the same concept value for the sampled condition $\C(s_1)=\C(s_2)$, thus the target submix would become either $\st=\mix$ or $\st=\bm{0}$. Except in Section \ref{sec:results:hard_negatives} where we provide an in-depth analysis of how the prior of these degenerate examples alters the operating point of our model, we always assume that during training the sampled sources can be separated using the concept $\ct$. All source audio files have a $4$-second length and are downsampled to $8$ kHz.

\subsection{Experimental details}
\label{sec:exp_framework:model}
\noindent{\bf Model:} We use a conditional Sudo rm -rf model (see Section \ref{sec:method:training}) with $B=16$ processing blocks. For efficient processing, we set the encoder and decoder filter lengths to $41$ taps with a hop size of $20$ time-domain samples. The number of learnable encoder/decoder bases and the number of intermediate channels are set to $512$. For all other parameters, we use the default values from the unconditional Sudo rm -rf model \cite{tzinis2022compute}. This setup leads to only a slight increase in trainable parameters compared to the unconditional version, from $9.66 \rightarrow 9.84$ millions.

\noindent{\bf Training:} The conditional models minimize the $l_1$ loss defined in Eq.~\eqref{eq:loss} while the unconditional models are optimizing the conventional PIT \cite{hershey2016deepclustering,Isik2016Interspeech09,Yu2017PIT} $l_1$ loss after computing the optimal assignment of the estimates to the target sources. We train all of the models using a batch size of $6$ using the Adam \cite{adam} optimizer with an initial learning rate of $10^{-3}$, halving it every $20$ epochs. We empirically found that $120$ epochs is enough training time for both PIT and conditional models to converge.

\noindent{\bf Evaluation:} We evaluate the source reconstruction fidelity of all the models at $120$ epochs using the median scale-invariant signal to distortion ratio (SI-SDR) \cite{leroux2019sdr} between the estimate $\est$ and the ground-truth target $\st$, as shown below:
\begin{equation}
\label{eq:sisdr}
    \begin{gathered}
    \operatorname{SI-SDR}(\est, \st) 
    = - 20 \log_{10} ( \nicefrac{\| \alpha \st \|}{\| \alpha \st  - \est\|} ),
    \end{gathered}
\end{equation}
where $\alpha = \st^\top \est /\| \es \|^2$ is a scalar which makes the metric invariant to scaling. The oracle PIT models are evaluated with SI-SDR but using the best possible assignment of the estimated sources to the target waveform. For the cases where the target submix is zero, $\st = \bm{0}$, thus, $\so = \mix$, we report the non-trivial $\operatorname{SI-SDR}(\eso, \so)$ which equivalently measures the reconstruction of the mixture at the non-target slot and is a meaningful metric since we enforce mixture consistency at the output of our models, namely $\mix = \est + \eso$. We want to underline that although all of our experiments display a similar behavior under the mean aggregation, we choose to report the median because of the susceptibility of SI-SDR to outliers. This metric instability becomes even more evident for conditional models that need to perform combined source ordering and separation.    

\section{Results and Discussion}
\label{sec:results}

\subsection{Cross-domain multi-conditioning}
\label{sec:results:multi_domain_cond}
Table~\ref{table:multi_condition} compares the proposed heterogeneous framework with a system using PIT training and oracle speaker selection on the SLIB and SVOX datasets. For the \textit{specialist} systems (annotated with $^*$), which are trained to focus on a single signal characteristic, the \textit{conditioned} models outperform their PIT counterparts with oracle selection for all conditions except SVOX language, likely because of the difficulty of simultaneously learning a harder concept like language and separating under challenging recording conditions.
We also note only a small drop in performance between the specialist and in-domain heterogeneous models, meaning a single model can learn to jointly perform targeted separation based on multiple signal characteristics. In addition, the most general multi-condition and multi-domain pre-trained network almost always outperforms unconditional separation with oracle source assignment (last 2 rows). Finally, heterogeneous training can aptly perform domain adaptation (see \textit{cross-domain heterogeneous} rows), since %
learning spatial conditioning appears to work well for SLIB when trained using spatial labels only for the SVOX data, and vice-versa. 
\begin{table}[t]
\scriptsize
\centering
  \sisetup{table-number-alignment = center,detect-weight=true,detect-inline-weight=math}
\caption{Median SI-SDR (dB) results of our proposed heterogeneous conditioning method on gender ($\G$), spatial location ($\S$), and language ($\L$) on the SVOX and SLIB dataset test partitions, for various combinations of training domains ($\setD$) and conditions ($\setC$). %
We also show PIT unconditional models with oracle target selection. $^*$ indicates specialist systems which are each trained and tested on a given pair of domain and condition.}
\vspace{-.3cm}
\setlength{\tabcolsep}{2.5pt}
\resizebox{\linewidth}{!}
{%
\begin{tabular}[]{lcc
*{4}{S[table-column-width=2.5em,table-format=3]}
S[table-format=3.1,table-column-width=3em]S[table-format=2.1,table-column-width=3em]S[table-format=3.1,table-column-width=3em]S[table-format=1.1,table-column-width=2.8em]}
\toprule
& & & \multicolumn{4}{c}{Train condition priors (\%)} & \multicolumn{4}{c}{Test conditions}\\
\cmidrule(lr){4-7}\cmidrule(lr){8-11}
\multicolumn{1}{l}{\multirow{3}{*}{\makecell{Training\\method}}}
& & & \multicolumn{2}{c}{SLIB} & \multicolumn{2}{c}{SVOX} &\multicolumn{2}{c}{SLIB} & \multicolumn{2}{c}{SVOX}\\
\cmidrule(lr){4-5} \cmidrule(lr){6-7} \cmidrule(lr){8-9} \cmidrule(lr){10-11} & $| \setD |$ & $|\setC|$ &  $\mathcal{G}$ &  $\mathcal{S}$ & $\mathcal{L}$   & $\mathcal{S}$ & $\mathcal{G}$ &  $\mathcal{S}$ & $\mathcal{L}$  & $\mathcal{S}$ \\
\midrule
Conditioned$^*$
 & 1 & 1 & 100 & 100 & 100 & 100 & \bfseries 11.4 & \bfseries 11.2 & 2.5 & \bfseries 9.1 \\
\midrule 
PIT (Oracle)$^*$ 
& 1 & 1 & 100 & 100 & 100 & 100 & 11.0 & 10.7 & 4.6 & 7.5 \\
\midrule
\multirow{2}{\spaceintable}{In-domain heterogeneous}
 & \multirow{2}{*}{1} & \multirow{2}{*}{2} & 50 & 50 & & & 10.9 & 10.7 & -0.5 & 8.6 \\
 &  &  & & & 50 & 50 & -0.6 & 6.2 & 3.2 & 6.8 \\
 \midrule
 \multirow{2}{\spaceintable}{PIT (Oracle)}
 & \multirow{2}{*}{1} & \multirow{2}{*}{2} & 50 & 50 & & & 9.5 & 8.9 & \bfseries 5.6 & 6.8 \\
 &  & & & & 50 & 50 & 5.2 & 4.5 & 4.6 & 5.6 \\
 \midrule
\multirow{5}{\spaceintable}{Cross-domain heterogeneous}
 & \multirow{4}{*}{2} & \multirow{4}{*}{2} & & 50 & 25 & 25 & -1.4 & 9.2 & 4.3 & 8.2 \\
 &  &  & 25 & 25 & & 50 & 9.9 & 9.9 & -0.7 & 9.0 \\
  &  &  & 50 & & & 50 & 10.1 & 8.9 & -0.9 & 9.0 \\
 &  &  & & 50 & 50 & & -0.5 & 8.4 & 4.3 & 6.8 \\\cmidrule(l){2-11} 
 & 2 & 3 & 25 & 25 & 25 & 25 & 8.9 & 8.7 & 4.4 & 7.8 \\
 \midrule
 \multirow{1}{\spaceintable}{PIT (Oracle)}
 & 2 & 3 & 25 & 25 & 25 & 25 & 8.0 & 7.3 & 5.5 & 6.5 \\
\bottomrule
\end{tabular}
\label{table:multi_condition}}
\vspace{-.3cm}
\end{table}

\subsection{Learning to generalize with bridge conditioning}
\label{sec:results:ood_generalization}
Table~\ref{table:bridge_condition} further explores domain adaptation, but in more severe conditions, as WSJ is anechoic and SLIB is reverberant. The proposed heterogeneous training method achieves 7.1 dB SI-SDR for the gender conditioned mixtures on SLIB, even though no SLIB gender-conditioned mixtures were used during training, compared to 5.8 and 4.2 dB when seeing no in-domain SLIB training data at all. We ablate the benefit of the proposed approach across two dimensions. First, we investigate the importance of having a ``bridge'' condition (here, the signal-energy characteristic $\E$), which is a common signal characteristic used to train the model across domains.
Removing the bridge condition leads to a performance drop in SLIB gender conditioned cases from 7.1 to 5.5 dB. Second, we evaluate the importance of the bridge condition concept to be as discriminative as possible. %
If the signal-energy bridge condition includes ambiguous energy mixtures (e.g., sources mixed at $\approx 0$ dB which do not contain a clearly louder speaker), performance on SLIB's gender-conditioned test set drops from 7.1 to 6.2 dB.   

\begin{table}[]
\scriptsize
\centering
\sisetup{table-number-alignment = center,detect-weight=true,detect-inline-weight=math}
\caption{Test median SI-SDR (dB) results for gender ($\G$) and signal-energy ($\E$) conditioning on WSJ and SLIB datasets with different cross-domain training seen conditions. ``Exclude amb. $\E$ cases'' indicates exclusion of training with ambiguous cases in $\E$ conditions: when sampling an energy conditioning vector ($c=\one[\E(s_1)]$), we only mix sources with a discriminative energy gap using a random input SNR of $\U [1, 5]$dB. When this is not enforced, input SNR is sampled from $\U [0, 5]$dB.}
\vspace{-.3cm}
\setlength{\tabcolsep}{0.2em}
\resizebox{\linewidth}{!}
{%
\begin{tabular}[t]{l
*{4}{S[table-column-width=2.5em,table-format=3]}
*{4}{S[table-column-width=3.2em,table-format=3.1]}}
\toprule
& \multicolumn{4}{c}{Train condition priors (\%)} & \multicolumn{4}{c}{Test conditions}\\ 
\cmidrule(lr){2-5}\cmidrule(lr){6-9} 
\multicolumn{1}{l}{\multirow{3}{*}{\makecell{Training\\method}}} & \multicolumn{2}{c}{WSJ} & \multicolumn{2}{c}{SLIB} &\multicolumn{2}{c}{WSJ} & \multicolumn{2}{c}{SLIB} \\
\cmidrule(lr){2-3} \cmidrule(lr){4-5} \cmidrule(lr){6-7} \cmidrule(lr){8-9}  & $\mathcal{G}$ & $\mathcal{E}$ & $\mathcal{G}$ & $\mathcal{E}$  & $\mathcal{G}$ & $\mathcal{E}$ & $\mathcal{G}$ & $\mathcal{E}$ \\
\midrule
\multirow{1}{2.5cm}{Proposed} & 25 & 25 & & 50 & 13.3 & 12.4 & 7.1 & 8.8 \\ \midrule
\multirow{1}{2.5cm}{(-) Bridge condition} & 50 & & & 50 & 14.5 & 7.4 & 5.5 & 9.2 \\ \midrule
\multirow{1}{2.5cm}{(-) Exclude amb.\ $\E$ cases} & 25 & 25 & & 50 & 13.0 & 11.8 & 6.2 & 8.4 \\ \midrule
\multirow{2}{2.5cm}{(-) In-domain data} & 100 & & & & \bfseries 17.3 & -2.4 & 5.8 & -2.3 \\
& 50 & 50 & & & 15.2 & \bfseries 14.3 & 4.2 & 3.0 \\ \midrule
\multirow{1}{*}{PIT (Oracle)*}& 100 & 100 & 100 & 100 & \bfseries 17.3 & 13.6 & \bfseries 10.9 & \bfseries 10.2 \\
\midrule
\multirow{1}{*}{PIT (Oracle)} &  25 & 25 & 25 & 25 & 12.9 & 11.9 & 9.3 & 8.5 \\
\bottomrule
\end{tabular}
\label{table:bridge_condition}
}
\vspace{-.4cm}
\end{table}

\subsection{Learning from degenerate conditions}
\label{sec:results:hard_negatives}
\begin{figure}[t]
  \centering
  \includegraphics[trim={0.7cm 0.75cm 0.65cm 0.75cm}, clip,width=1.\linewidth]{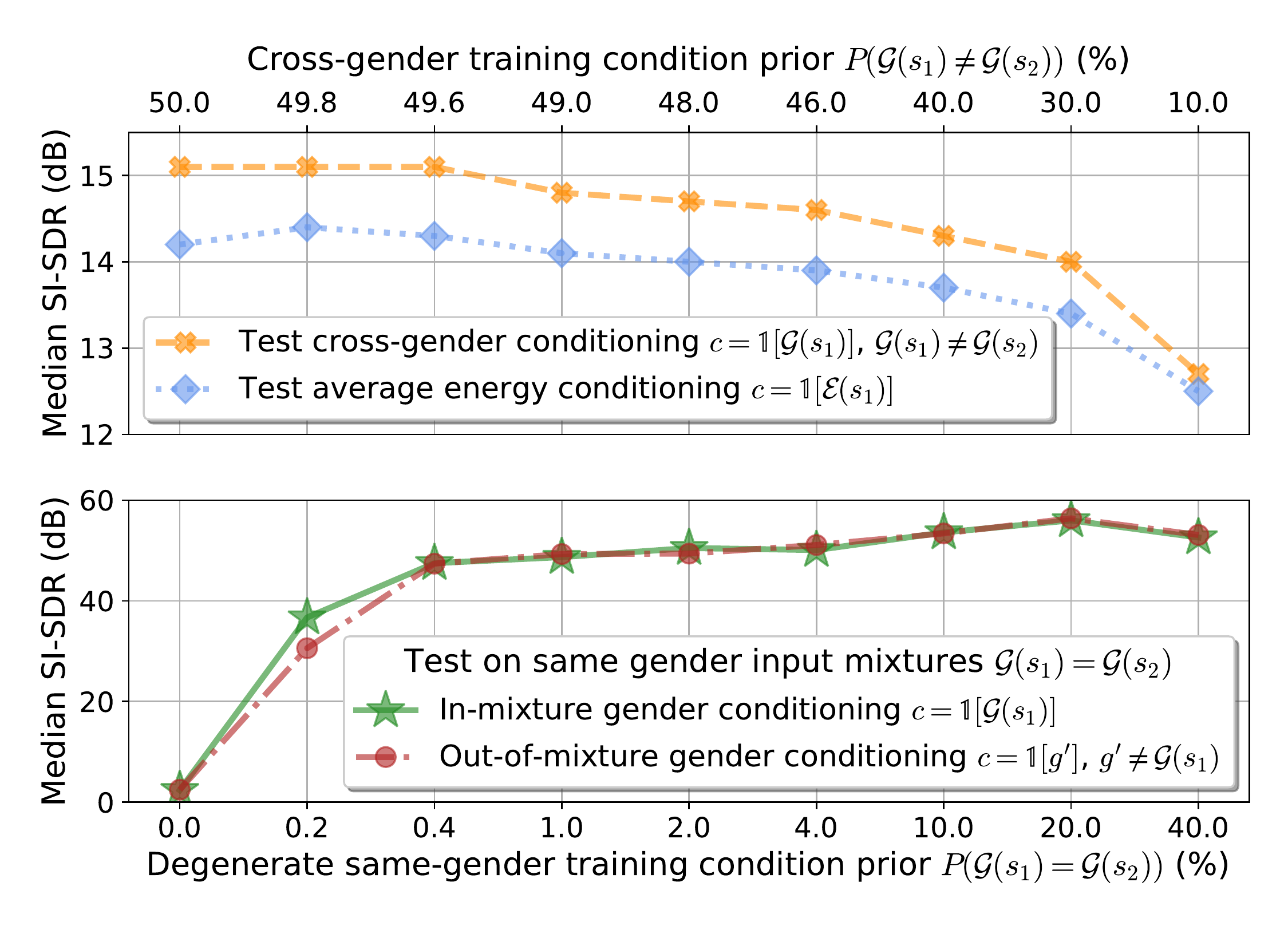}
\vspace{-.5cm}
\caption{Median SI-SDR performance on the WSJ test set for various conditions when training with a static prior of sampling an energy condition vector $P(c=\one(\E(s_1)))=50\%$ and sweeping through the percentage of degenerate gender conditioning. In essence, the leftmost points correspond to training a model with only easy cross-gender mixtures whereas as we move towards the rightmost points, the percentage of the degenerate gender conditioning with same-gender inputs increases.}
\label{fig:wsj_abl}
\vspace{-.2cm}
\end{figure}

A heterogeneous target speech separation model allows a user to select the signal characteristic most relevant for isolating the desired speaker in a specific mixture. However, to make a truly robust system, our model should still behave as expected when the query is a degenerate, meaning that the target for a given query is a zero waveform $\st=\bm{0}$ (e.g., we condition on ``French'' and the mixture contains only ``English'' speakers), or when the query matches all sources in the mixture itself. To do so, we control the prior distribution for a concept $P(v)$ to be non-discriminative, where $C(s_1)=C(s_2)$. To the best of our knowledge, shifting the operating point for controlling the robustness of separation systems has only been studied for unconditional separation models \cite{tzinis2021unified} and a few works have considered degenerate conditions in target speech extraction but without thoroughly investigating the influence on performance of the ratio of such conditions in the training data \cite{zhang2020x,borsdorf21_interspeech}.
Because separating degenerate conditions as a task by itself is trivial, i.e., the target is either silence or the mixture, such samples can easily saturate the loss function during training. Figure \ref{fig:wsj_abl} shows the trade-off between the percentage of degenerate gender condition cases used during training with WSJ and the performance on various conditioning tasks on WSJ test sets. We evaluate the performance on cases where the concept is discriminative (top-panel) and non-discriminative (bottom panel). We see that including no degenerate causes the system to fail on same gender mixtures, but including even 1\% degenerate examples during training degrades the performance for cross-gender mixtures. For this dataset, there appears to be a sweet-spot when training with 0.4\% degenerate condition examples. 

\subsection{Robust learning of new concepts using more conditions}
\label{sec:results:harder_conditions}
\begin{figure}[]
  \centering
  \includegraphics[trim={0.6cm .8cm 0.65cm 0.7cm}, clip,width=1.\linewidth]{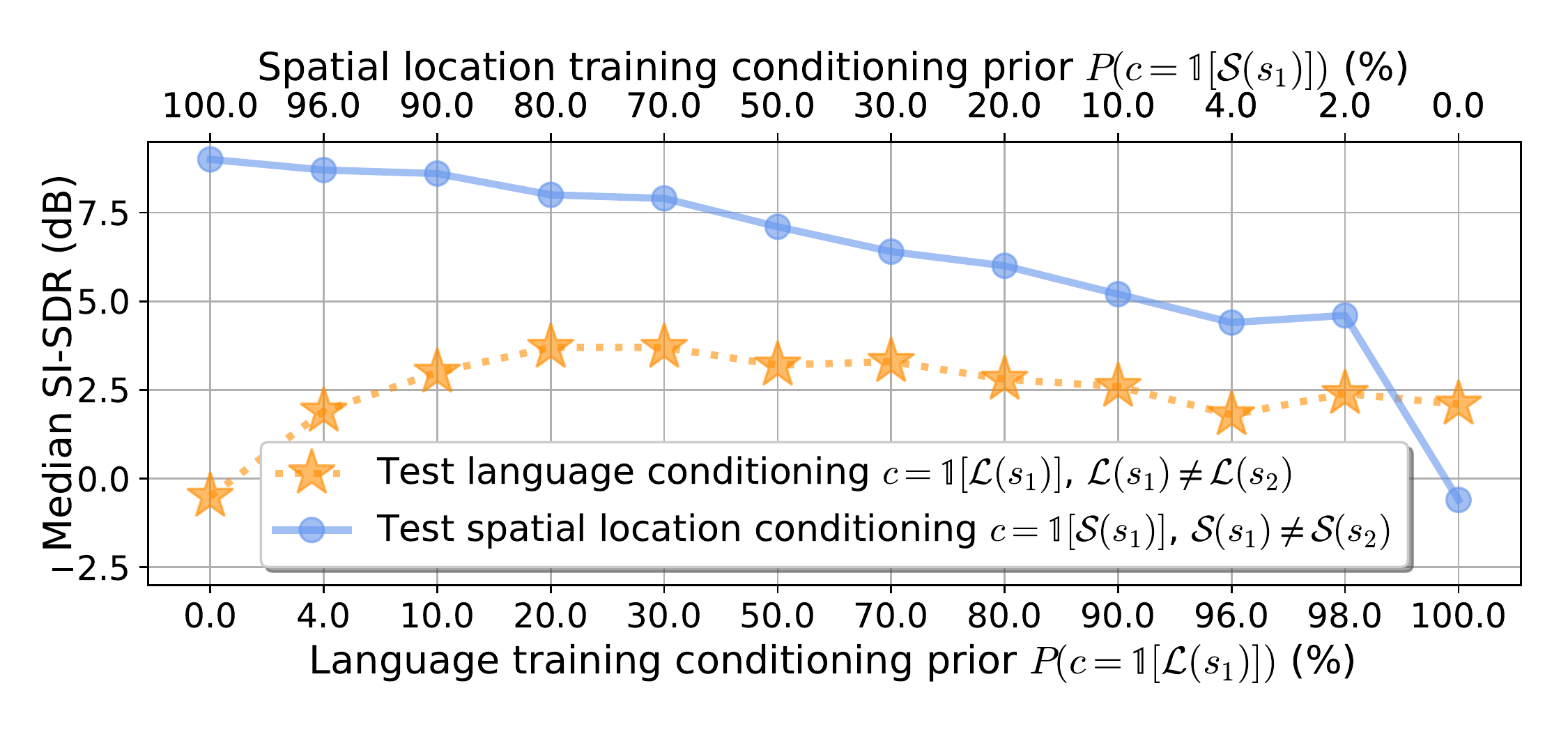}
  \vspace{-.5cm}
\caption{Evaluation of the median SI-SDR performance on the SVOX test set for language and spatial conditioning when training with different priors of the input conditional vectors. From left to right we increase the amount of the harder language conditioning $P(c=\one(\L(s_1))): 0 \rightarrow 1$ and (inversely $P(c=\one(\S(s_1))): 1 \rightarrow 0$ for the complementary easier spatial conditioning). Note that the best model for the harder language condition is not the one trained solely for this task.}
\label{fig:gsvox_abl}
\vspace{-.5cm}
\end{figure}

As shown in Table~\ref{table:multi_condition}, SVOX conditioned on language has the lowest SI-SDR scores, possibly because we have four possible languages, thus, there is a higher prior probability of picking a speaker with the incorrect language. In Fig.~\ref{fig:gsvox_abl}, we explore how adding an easier signal characteristic, spatial location $\S$, improves performance on the more difficult language conditioning $\L$. Surprisingly, the best performance for language conditioning is achieved when only 20\% of the training mixtures are conditioned using $\L$ and 80\% are conditioned using $\S$. Unfortunately, the inverse relationship does not hold, i.e., including the more difficult condition $\L$ negatively impacts performance for the easier condition $\S$.  We suspect this is due to model parameters that are unrelated to the conditioning, such as the learned encoder and decoder receiving large gradient updates for cases when the incorrect language is chosen, which may cause them to be less effective for speech separation in general. %
Mitigating this issue is an important topic of future work. 

\section{Conclusion}
\label{sec:conclusion}
In this paper, we have introduced the heterogeneous target source separation task based on several non-mutually exclusive signal characteristic conditions. We have experimentally shown that the proposed heterogeneous condition training framework provides benefits in terms of domain generalization and has the ability to leverage ``easy'' conditions to facilitate training with more difficult concepts. Heterogeneous conditioning can be made robust to degenerate cases and often performs better than unconditional models with oracle source assignment. In the future, we plan to extend our method to incorporate a variable number of speakers and training with self-supervised schemes.

\balance
\bibliographystyle{IEEEtran}
\bibliography{refs}

\end{document}